\documentclass[twocolumn,showpacs,preprintnumbers,amsmath,amssymb]{revtex4}
\usepackage{graphicx}
\usepackage{dcolumn}
\usepackage{bm}

\newcommand{\C}{\ensuremath{\mathbb{C}}}

\newcommand{\Z}{\ensuremath{\mathbb{Z}}}

\newcommand{\R}{\ensuremath{\mathbb{R}}}
\newcommand{\bR}{\overline{\R}}
\newcommand{\bRp}{\overline{\R_+}}
\renewcommand{\S}{\mathbb{S}}
\newcommand{\K}{\mathcal K}
\newcommand{\B}{\mathcal B}

\newcommand{\J}{\mathcal J}
\newcommand{\E}{\mathcal E}

\renewcommand{\H}{\mathcal H}
\newcommand{\ind}{\mbox{ind}}
\newcommand{\Ind}{\mbox{index}}

\newcommand{\tr}{\mbox{tr}}
\newcommand{\Tr}{\mbox{Tr}}
\renewcommand{\d}{{\mathrm d}}
\newcommand{\Om}{\Omega}
\newcommand{\Omt}{\tilde{\Omega}}

\newcommand{\x}{x}

\def\ev{e}
\def\odd{o}

\begin{document}

\title{The topological meaning of Levinson's theorem, half-bound states included}

\author{Johannes Kellendonk}
\author{Serge Richard}
\affiliation{    \begin{itemize}
    \item[]
Institut Camille Jordan, CNRS UMR 5208,
Universit\'e Lyon 1, Universit\'e de Lyon,
43 boulevard du 11 novembre 1918, F-69622 Villeurbanne cedex,
France
    \item[]
      \emph{E-mails\:\!:}
      kellendonk@math.univ-lyon1.fr\,~and\,~srichard@math.univ-lyon1.fr
    \end{itemize}}

\date{\today}

\begin{abstract}
We propose to interpret Levinson's theorem as an index theorem.
This exhibits its topological nature. It furthermore leads to a more
coherent explanation of the corrections due to resonances at
thresholds.
\end{abstract}

\maketitle
\section{Introduction}

Levinson's theorem is a relation between the number of bound states
of a system and an expression related to the scattering part of that
system. The latter expression can be written either in terms of an
integral over the time delay, or as an evaluation of the spectral
shift function (see the review papers \cite{Bolle,Nuss}).
In the simplest situations, the relation is an
equality, but that is not always the case. Depending on the space
dimension and on the existence of resonances at thresholds, also
called half-bound states, corrections to the former equality have to
be taken into account.

We propose here a topological approach to this theorem by
interpreting it as an index theorem. This does not only shed new
light on it, but it provides also a more coherent and natural way to
take the corrections into account.
It is inspired by Bellissard's approach to topological
phenomena in solid states physics \cite{Bel} and was first proposed
for simpler models in \cite{KROAMP}.
The proof relies on the use of  boundary maps
between the topological invariants of the bound part and of the
continuous part of the system. It puts emphasis on the wave
operators.

The content of this Letter is the following: First we recall a
common form of Levinson's theorem written in terms of the time
delay. In Sections \ref{sec2} and \ref{sec3}, we expose our
topological approach and introduce the framework suitable for
Schr\"odinger operators on $\R^n$. In the last section, we
illustrate our ideas with systems on $\R$.

\section{A common form of Levinson's theorem}

We consider a quantum mechanical system described by a Hamiltonian
$H$ in a Hilbert space $\H$. The spectrum $\sigma(H)$ of $H$ is
composed of eigenvalues and of a continuous part. The eigenfunctions
of $H$ span a subspace $\H_p$ of $\H$, and the projection onto
$\H_p$ is denoted by $P_p$.

The continuous part of the system can generally be described by
scattering theory, that is, by comparison of the dynamics generated
by $H$ with the dynamics generated by a simpler Hamiltonian $H_0$.
The latter operator is assumed to be absolutely continuous. We also
assume that the following strong limits exist:
\begin{equation*}
\Omega_\pm = \mbox{s}\! - \!\lim_{t\to\pm\infty}e^{itH} e^{-itH_0}\
,
\end{equation*}
and that the so-defined wave operators are asymptotically complete.
It follows that the scattering operator ${S} = \Omega_+^*\Omega_-$
is unitary and that the isometry $\Omega:=\Omega_-$ has support and
range projections
\begin{equation}\label{eq-om}
\Om^*\Om = 1,\qquad \Om \Om^*= 1-{P_p} \ .
\end{equation}
Furthermore, since $S$ commutes with $H_0$, it is unitarily
equivalent to an operator-valued function $\sigma(H_0)\ni \lambda
\mapsto S(\lambda)\in \B(\H_\lambda)$ resulting from the direct
integral decomposition of $\H=\int_{\sigma(H_0)} \H_\lambda \d
\lambda$ with respect to $H_0$. The operator $S(\lambda)$ is
referred to as the $S$-matrix at energy $\lambda$, and
$iS^*(\lambda) S'(\lambda)$ as the time delay operator at
energy $\lambda$. Then, a common form of Levinson's theorem is:
\begin{equation} \label{eq-lev}
\frac{1}{2\pi}\int_{\sigma(H_0)} \big( \tr_\lambda
[iS^*(\lambda)S'(\lambda)] - c(\lambda)\big) \d \lambda =
\mbox{Tr}(P_p) + \nu\ .
\end{equation}
Here $\tr_\lambda$ is the operator trace on $\H_\lambda$ and
$\mbox{Tr}$ is the operator trace on $\H$. In particular
$\mbox{Tr}(P_p)$ counts the number of bound states. The regularizing
term $c(\lambda)$ is necessary if the map $\lambda\mapsto
\tr_\lambda [iS^*(\lambda) S'(\lambda)]$ is not integrable. The
correction term $\nu$ arises from the existence of resonances at
thresholds in the spectrum of $H$.
The explanation for the presence of $\nu$ in \eqref{eq-lev} is sometimes quite ad hoc.

For example, for Schr\"odinger operators on $\R^n$, the correction
depends on the existence of $0$-energy resonances and on the
dimension $n$. $0$-energy resonances are solutions of the equation
$H\Psi = 0$ with $\Psi$ not in $L^2(\R^n)$ but in some suitable larger space.
If $n=3$ and if such a $0$-energy resonance exists, the correction
$\nu$ is equal to $1/2$. In other dimensions the picture is
different.

\section{Topological approach}\label{sec2}

In this section, we show how to rewrite \eqref{eq-lev} as an index
theorem. Our approach is based on the following
construction: Let $\E$ be a closed unital subalgebra of $\B(\H)$
containing an ideal $\J$. Let us assume that (i) $\Om$ belongs to
$\E$, (ii) the image of $\Om$ through the quotient map $q: \E\to
\E/\J$ is a unitary operator incorporating $S$. We shall see that in
the simplest situation, $q(\Om)=S$, but that in the general case,
$q(\Om)$ incorporates besides $S$ other components which account for
the correction in \eqref{eq-lev}.

Before explaining how to construct $\E$ we demonstrate how these
assumptions lead to a topological version of Levinson's theorem. We
think of $\J$ as the algebra related to the bound states system,
and of $\E/\J$ as the one corresponding to the scattering system. By
the general machinery of $K$-theory of $C^*$-algebras the map $q$
gives rise to a topological boundary map, called the index map,
$\ind:K_1(\E/\J)\to K_0(\J)$
which can be described as follows: If $u\in \E/\J$ is a unitary
representing an element $[u]_1$ in $K_1(\E/\J)$ and having a
preimage $\omega \in \E$ under $q$ which is an isometry, then
$\ind([u]_1) = [\omega \omega^*]_0-[\omega^* \omega]_0$, the
difference of the classes in $K_0(\J)$ of the range and the support
projections of $\omega$. Thus if $\Om$ belongs to $\E$, the relation
\eqref{eq-om} yields
\begin{equation}\label{eq-lev-ab}
\ind([q(\Om)]_1) = - [P_p]_0\ .
\end{equation}
This result is our abstract Levinson's theorem. In the simplest
situation, this relation reads $\ind([S]_1) = - [P_p]_0$, but in the
general case the corrections arise from the difference between
$[q(\Omega)]_1$ and $[S]_1$. Concrete Levinson theorems like
\eqref{eq-lev} arise if we apply functionals to the $K$-groups to
obtain numbers.

For a large class of scattering systems including potential
scattering we expect that $\J= \K(\H)$, where $\K(\H)$ is the
algebra of compact operators on $\H$, and that $\E/\J$ is isomorphic
to $C(S^1,\K(\H_\lambda))$, the continuous functions on the circle
with values in $\K(\H_\lambda)$. In that case $K_0(\J)$ and
$K_1(\E/\J)$ are both isomorphic to $\Z$ so that, up to a
normalization, the only functional on $K_0(\J)$ is given by the
trace $\Tr: [p]_0\mapsto \Tr(p)$, and the only functional on
$K_1(\E/\J)$ is given by $w: [u]_1\mapsto w(u)$, the winding number of
$t\mapsto \det\big(u(t)\big)$, the
determinant possibly needing regularization.
In particular
\eqref{eq-lev-ab} reduces to the index theorem of Krein-Gohberg
\begin{equation}\label{eq-lev2}
w(q(\Om)) = \Ind(\Om) = -\Tr(P_p)\ .
\end{equation}
This is our formulation of the concrete Levinson's theorem
\eqref{eq-lev}. Note that there is room for further, potentially
unknown, identities of Levinson type by choosing other functionals
in cases in which the $K$-groups are richer than those considered
above.

\section{Constructing the algebras}\label{sec3}

We do not expect that there is a universal construction of the
algebra $\E$ and its ideal $\J$, rather to the contrary, we believe
that part of the richness of the theory lies in the flexibility of
their choice. But for simple scattering systems, {\it i.e.}~when
$H_0$ is the free Laplacian $-\Delta$ on $\R^n$,
we construct $\E$ with the help of the conjugate operator to $H_0$, namely the generator
$A$ of dilations. In $\H=L^2(\R^n)$, the operators $H_0$ and $A$ have a purely absolutely
continuous spectrum equal to $\R_+$ and $\R$, respectively.

Let $\E'$ be the closure in $\B(\H)$ of the algebra generated by
elements of the form $\eta(A)\psi(H_0)$, where $\eta$ is a
continuous function on $\R$ which converges at $\pm \infty$, and
$\psi$ is a continuous function $\R_+$ which converges at $0$ and
at $+\infty$. Stated differently, $\eta\in C(\bR)$, where
$\bR=[-\infty,+\infty]$ is the two point compactification of $\R$,
and by analogy $\psi \in C(\bRp)$. Let $\J'$ be the norm closed
algebra generated by $\eta(A)\psi(H_0)$ with $\eta\in C_0(\R)$ and
$\psi \in C_0(\R_+)$ {\it i.e.}~all limits vanish. $\J'$ is
an ideal of $\E'$.

To describe the quotient $\E'/\J'$ we consider the square
$\square:=\bRp\times \bR$ whose boundary $\partial (\square)$ is the
union of four parts: $\partial (\square) =B_1\cup B_2\cup B_3\cup
B_4$, with $B_1 = \{0\}\times \bR$, $B_2 = \bRp \times \{+\infty\}$,
$B_3 = \{+\infty\}\times \bR$ and $B_4 = \bRp\times \{-\infty\}$. We
can also identify $C(\partial (\square))$ with the subalgebra of
$C(\bR)\oplus C(\bRp)\oplus C(\bR)\oplus C(\bRp)$ given by elements
$(\Gamma_1,\Gamma_2,\Gamma_3,\Gamma_4)$ which coincide at the
corresponding end points, that is, for instance,
$\Gamma_1(+\infty)=\Gamma_2(0)$. Then, the quotient map $q':\E'\to
\E'/\J'\cong C(\partial(\square))$, evaluated on elements generating
$\E'$, is given by $q'\big(\eta(A)\psi(H_0)\big)= \big(\Gamma_1(A),
\Gamma_2(H_0), \Gamma_3(A), \Gamma_4(H_0)\big)$, where
$\Gamma_1(A) = \eta(A)\psi(0)$, $\Gamma_{2}(H_0) =
\eta(+\infty)\psi(H_0)$,
$\Gamma_{3}(A) = \eta(A)\psi(+\infty)$ and $\Gamma_{4}(B) =
\eta(-\infty)\psi(H_0)$.
Observe that $\mbox{s}\!-\!\lim_{t\to
\pm\infty}e^{itB}\eta(A)\psi(H_0)e^{-itB}$, with $B=\frac{1}{2}\ln
(H_0)$, are equal to $\Gamma_2(H_0)$ and $\Gamma_4(H_0)$,
respectively.

Finally, $\E$ and $\J$ are obtained by adding to the generators of
$\E'$ and $\J'$ all compact operators in angular momentum. The unit
$1$ is also added to $\E$. Since $H_0$ and $A$ are rotation
invariant, such modifications do not perturb the above picture. In
particular, $\J=\K(\H)$ and $\E/\J \cong C\big(\partial(\square),\C
+ \K[L^2(\S^{n-1})]\big)$.

Our basic assumption is that $\Om$ belongs to $\E$. From the above
observation, the intertwining relation and the invariance principle,
it follows that $\Gamma_2(H_0)= S$ and $\Gamma_4(H_0) = 1$.
The winding number $w(q(\Om))$ is the sum of four terms, each side
of the square contributing for one.  In regular cases,
the winding along $B_j$ contributes with $w_j
= \frac{1}{2\pi i} \int_{B_j} \tr[\Gamma_j^*\;\!\d \Gamma_j]$.
Then $w_2 = \frac{1}{2\pi i} \int_0^\infty \tr_\lambda
[S^*(\lambda) S'(\lambda)]\d \lambda$
and $w_4 = 0$.
Comparing
\eqref{eq-lev} with \eqref{eq-lev2} we see therefore that the
correction term arises now on the l.h.s.~of the equality from the
possible contribution of $\Gamma_1$ and $\Gamma_3$ to the winding
number. If $c(\lambda)\neq 0$ the above formulas have to be regularized.

We finally note that \eqref{eq-lev2} can be refined: if $P$ is a
projection in $\J$ which commutes with $\Omega$ then $\ind[q(\Om P)
]_1 = - [P_p P]_0$ leading to $ w(q(\Om P)) = \Ind(\Om P)$. For
example, choosing for $P$ a projection on an angular momentum sector
leads to a Levinson's theorem for that sector.

\section{One dimensional scattering}

We illustrate our approach with one-dimensional systems described by
Schr\"odinger operators on $\H=L^2(\R)$, first with
$-\Delta$ perturbed by a one point interaction, and second with $-\Delta$
perturbed by multiplication operators. In both cases,
$\H_\lambda=\C^2$ and $\E/\J \cong
C\big(\partial(\square),M_2(\C)\big)$. Our aim is to obtain a
formula for $\Om$ which shows that it belongs to $\E$, to determine
each $\Gamma_j$ and to show how they contribute to $w(q(\Om))$. For
that purpose, the following observation taken from \cite{KR06} is
essential: Let $g$ be a smooth rapidly decreasing
function on $\R$ and $T$ be the operator defined
by $[Tg](r\omega) =
\hbox{$\frac{1}{\sqrt{2\pi}}$}\int_0^\infty e^{i\kappa
r}\hat{g}(\kappa \omega)\d \kappa$, with $r\geq 0$,
$\omega\in\{+1,-1\}$ and $\hat{g}$ the Fourier transform of $g$. 
Then $T$ extends to
the operator $\hbox{$\frac{1}{2}$}\big(1-R\big)$ with
\begin{equation*}
R:=R(A) = r_{\ev}(A)P_\ev + r_{\odd}(A) P_{\odd}\ ,
\end{equation*}
where $P_\ev, P_{\odd}$ are the projections onto the
even (symmetric), odd functions of $\H$, respectively, and
\begin{equation*}
r_{\ev}(x)  := -\tanh(\pi x) - i[\cosh(\pi x)]^{-1}\ , \quad
r_o := \overline{r_e} \ .
\end{equation*}
Clearly, $r_\ev$ and $r_\odd$ belong to $C(\bR)$.

\subsection{One dimensional point interactions}

Schr\"odinger operators with one point interaction at the origin can
be defined as the family of self-adjoint extensions of the
restriction of the Laplacian on a suitable subset of $L^2(\R)$
\cite{AGHH}. Different point interactions arise from different
extensions and we concentrate here on the two families of point
interaction called $\delta$-interaction and $\delta'$-interaction.
In these cases  the wave operator has the form \cite{KR06}~:
\begin{equation*}
\Omega = 1+\hbox{$\frac{1}{2}$}(1-R)\;\!(S-1)\ .
\end{equation*}
Let us stress that the first factor is universal and does not depend
on the choice of any self-adjoint extension, only the $S$-term
depends on such a particular choice. We shall see later that a
similar form holds for the wave operator in the case of potential
scattering. Nevertheless, the contributions to the winding number
corresponding to $\Gamma_1$ and $\Gamma_3$ clearly depend on the
different behaviour of the matrix $S(\lambda)$ for $\lambda =0$ or
$\lambda = +\infty$. For example if $S(0)=1$, then $\Gamma_1=1$.
More interesting phenomena arise if $P_\ev S(0)\neq P_\ev$ or
$P_\odd S(0)\neq P_\odd$, as exhibited in the following situations.

The family of extensions called $\delta$-interaction is
parameterized by $\alpha\in \R\cup\{\infty\}$. The parameter
describes the boundary condition of the wave function
$\Psi'(0_+)-\Psi'(0_-) = \alpha \Psi(0)$ which can be formally
interpreted as arising from a potential $V=\alpha\delta$ where
$\delta$ is the Dirac $\delta$-function at $0$. The extension for
$\alpha=0$ is equal to $H_0$ and the extension for $\alpha=\infty$
is the Laplacian with a Dirichlet boundary conditions at the origin. These
extensions have a single eigenvalue if $\alpha<0$ and no eigenvalue
if $\alpha \in [0,\infty]$. The scattering operator is given by
$S = s^\alpha(H_0) P_{\ev} + P_{\odd}$
with $s^\alpha(\lambda) = \frac{2\sqrt{\lambda}-i\alpha}{2\sqrt{\lambda}+i\alpha}$.
Note that $s^\alpha \in C(\bRp)$ with values at $0$ and $+\infty$ depending
on $\alpha$.

The family of extensions referred to as $\delta'$-interaction is
parameterized by $\beta\in\R\cup\{\infty\}$, the parameter describing
the boundary condition of the wave function $\Psi(0_+)-\Psi(0_-) =
\beta \Psi'(0)$. This can be formally interpreted as arising from a
potential $V=\beta\delta'$. The extension for $\beta=0$ is equal to
$H_0$ and the extension $\beta=\infty$ is the Laplacian on $\R$ with
Neumann boundary conditions at the origin. These extensions possess a single
eigenvalue if $\beta < 0$ and no eigenvalue if $\beta \in
[0,\infty]$. The scattering operator is
$S =  P_{\ev} + s^\beta(H_0)  P_{\odd}$
with $s^\beta(\lambda) =
\frac{2 + i \beta \sqrt{\lambda}}{2 - i \beta \sqrt{\lambda}}$.
Again $s^\beta \in C(\bRp)$ with values at
$0$ and $+\infty$ depending on $\beta$.

In all these examples, the wave operator $\Om$ clearly belongs to
the algebra $\E$ introduced above.  Since $\Om$ commutes with
$P_\ev$ and $P_\odd$ we obtain a Levinson's theorem for each sector
separately. But for $\delta$-interactions $\Om_\odd:=\Om P_\odd = P_\odd$ and
hence the odd sector theorem is trivial. Likewise the even sector theorem is trivial for a
$\delta'$-interaction. We present the non-trivial results in the two tables below
with the notations
$\Gamma_i^{o/e}:=\Gamma_iP_{o/e}$ and $w_i^{o/e}:=w(\Gamma_i^{o/e})$.
\begin{center}
\begin{tabular}{|c|c|c|c|c|c|c|c|c|c|}
\hline $\delta$-interaction&
$\Gamma_1^e$ & $\Gamma_2^e$ & $\Gamma_3^e$
& $\Gamma_4^e$
& $w_1^e$ & $w_2^e$ & $w_3^e$ & $w_4^e$ & $w(q(\Om_e))$ \\
\hline\hline $\alpha < 0$ &$ r_{\ev} $&$ s^\alpha $&$ 1 $&$ 1 $&$
-\frac{1}{2} $&$ -\frac{1}{2}  $&$ 0 $&$0 $&$ -1 $ \\\hline $\alpha
= 0  $&$  1 $&$ 1 $&$ 1 $&$ 1 $&$ 0  $&$0$& $ 0$&$0 $&$ 0 $ \\\hline
$\alpha > 0  $&$ r_{\ev} $&$ s^\alpha $&$ 1 $&$ 1 $&$ -\frac{1}{2}
$&$ \frac{1}{2}  $&$ 0 $&$0 $&$ 0 $ \\ \hline $\alpha = \infty  $&$
r_{\ev} $&$ -1 $&$ r_{\ev} $&$ 1 $&$ -\frac{1}{2} $&$0$& $
\frac{1}{2} $&$0 $&$ 0 $ \\
\hline
\end{tabular}
\end{center}
\begin{center}
\begin{tabular}{|c|c|c|c|c|c|c|c|c|c|}
\hline $\delta'$-interaction
& $\Gamma_1^o$ & $\Gamma_2^o$ & $\Gamma_3^o$
& $\Gamma_4^o$
& $w_1^o$ & $w_2^o$ & $w_3^o$ & $w_4^o$ & $w(q(\Om_o))$ \\
\hline\hline $\beta < 0$ &$ 1 $&$ s^\beta $&$ r_{\odd} $&$ 1 $&$ 0
$&$ -\frac{1}{2}  $&$ -\frac{1}{2}$&$0 $&$ -1 $ \\\hline
$\beta = 0  $&$ 1 $&$ 1 $&$ 1 $&$ 1 $&$ 0 $&$ 0 $&$ 0 $&$0 $&$ 0 $ \\
\hline
$\beta > 0  $&$ 1 $&$ s^\beta $&$ r_{\odd} $&$ 1 $&$ 0 $&$ \frac{1}{2}  $&$
-\frac{1}{2}$&$0 $&$ 0 $ \\
\hline
$\beta = \infty  $&$  r_{\odd} $&$ -1 $&$ r_{\odd} $&$ 1 $&$ \frac{1}{2}  $&$0$& $
-\frac{1}{2}$&$0 $&$ 0 $ \\\hline
\end{tabular}
\end{center}
We thus  see that both, $w_1$ and $w_3$ contribute to
the correction term $\nu$ in (\ref{eq-lev}).

\subsection{One dimensional potential scattering}

In this section, we consider Schr\"odinger operators of the form
$H=H_0+V$, with potential $V$ given by a multiplication
operator. If the potential is regular enough and vanishes
sufficiently rapidly at infinity,
the wave operator $\Om$ can be expressed with the help of the solution $\Psi$ of the
Lippmann-Schwinger equation:
\begin{equation*}
[\Om g](x)=\hbox{$\frac{1}{\sqrt{2\pi}}$} \int_\R \Psi(k,x)\;\!
\hat{g}(k)\;\!\d k\ ,
\end{equation*}
with $g$ as above.
The solution has asymptotic behaviour
\begin{equation}\label{asympt}
\Psi (k,x)\stackrel{|x|\to\infty}{\sim} e^{i k\cdot x} +e^{i \kappa
r} f(\kappa^2,\omega_k,\omega_x)\ ,
\end{equation}
where $k = \kappa \omega_k$,
$x=r\omega_x$, and $f$ is the scattering amplitude.
Furthermore, the coefficients of the scattering matrix $S(\lambda)$
at energy $\lambda = \kappa^2$ in the momentum representation
are given by
$1+f(\kappa^2,\pm1,\pm1)$.

Let us now consider the integral operator $\Omt$ defined by
\begin{eqnarray*}\label{eq-op}
[\Omt g] (\x) &=& \hbox{$\frac{1}{\sqrt{2\pi}}$}\int_{\R} e^{i
\kappa r}
\;\!f(\kappa^2,\omega_k,\omega_x)\;\! \hat{g}(k) \;\!\d k \\
&=&\hbox{$\frac{1}{\sqrt{2\pi}}$}\int_{\R_+} e^{i \kappa
r}\big[\big(S(\kappa^2)-1\big)\hat{g}\big](\kappa\omega_x) \;\!\d
\kappa \\
&=& \big[\hbox{$\frac{1}{2}$} (1 - R)(S - 1)g\big](x)\ .
\end{eqnarray*}
Then, it follows that
\begin{equation*}
\Om= 1+ \hbox{$\frac{1}{2}$} (1 - R)(S - 1) + K\ ,
\end{equation*}
with
$[Kg](x) = \hbox{$\frac{1}{\sqrt{2\pi}}$} \int_\R \rho(k,x)\;\!
\hat{g}(k)\;\!\d k$, $\rho(k,x)$ being the
remainder in the asymptotic expansion \eqref{asympt}. In particular
$\Om$ belongs to $\E$ provided the map $\R_+ \ni \lambda \mapsto
S(\lambda)\in M_2(\C)$ is continuous and has limits at $0$ and
$+\infty$, and $K$ is compact. Both conditions require further assumptions
on the potential which go beyond the once implicitly
assumed for the validity of the above approach.
Without aiming at the most general case here, we can say the following:
The left ($x<0$) and the right part ($x>0$) of the remainder $\rho$
satisfy Jost type equations which can be solved by fixed point methods.
We find that $\rho$ is square integrable and hence $K$ compact provided
$|V(x)|\leq C(1+|x|)^{-\frac{5}{2}-\epsilon}$, $\epsilon>0$.
This condition is sufficient to conclude that $\Om$ belongs to $\E$.

We finally explain how the correction term $\nu$ of \eqref{eq-lev} arises in
our approach.
For that purpose we
use a basis for $M_2(\C)$ in which
$R=\left(\begin{array}{cc}
r_e & 0 \\
0 & r_o
\end{array}\right)$.
It corresponds to the decomposition of $L^2(\R)$
into even and odd sectors. The form of $S(0)$ falls into two cases,
characterized by the value of $\det(S(0))$. One finds accordingly \cite{AK}
\begin{equation}\label{S0}
S(0) = \left(\begin{array}{cc}
-1&0 \\
0&1
\end{array}\right)
\quad \hbox{or} \quad
\hbox{$\frac{1}{\gamma^2+1}$}\left(\begin{array}{cc}
2\gamma&1-\gamma^2 \\
\gamma^2-1&2\gamma
\end{array}\right)
\end{equation}
with $\gamma \in \R^*$. The first case occurs if $H$ does not
admit a resonance at energy zero, it is referred to as the
generic case (g.c.). The second, so-called exceptional case (e.c.),
occurs when such a zero energy resonance exists.
The contribution to the winding number coming from $\Gamma_1$
can be determined:
$w(\Gamma_1)=-\hbox{$\frac{1}{2}$}$ in the generic case,
and $w(\Gamma_1)=0$ in the exceptional one.
Thus, taking into account that $\Gamma_3 = \Gamma_4=1$ (the former
because $S(\infty) = 1$) one obtains from \eqref{eq-lev2}
\begin{equation}\label{presquefini}
\frac{1}{2\pi}\int_{\R_+} \tr_\lambda
[i S^*(\lambda)S'(\lambda)]  \d \lambda =
\left\{\begin{array}{ll}
{N -\frac{1}{2}}, & \hbox{g.c.} \\
{N}, & \hbox{e.c.}
\end{array}\right.
\end{equation}
where $N= \hbox{Tr}(P_p)$ is the number of bound states of $H$.
In particular, the correction term $\nu$ corresponds to $w_1$.
This result is in accordance with the literature \cite{BGW,BGK,Ma,Sassoli}.

If the potential is symmetric, a Levinson's theorem holds for
each sector. In that situation, the exceptional case $\gamma=1$ in
\eqref{S0} corresponds to an even zero energy resonance, and
the case $\gamma=-1$ corresponds to an odd zero energy resonance.
The results for the even and odd sector are summarized in the
following two tables. \medskip

\begin{center}
\begin{tabular}{|c|c|c|c|c|c|c|}
\hline
even sector & $\Gamma_1^e$ & $\Gamma_2^e $ & $S_e(0)$ & $w_1^e$
& $w_2^e$ & $w(q(\Om_e))$ \\ \hline\hline
g.c.\  &$ r_{\ev} $&$ S_{\ev} $& $-1$ & $ -\frac{1}{2}  $&$
-(N_{\ev}-\frac{1}{2})$&$ -N_{\ev} $  \\\hline
e.c.\ &$1$&$ S_{\ev} $& $1$& $  0 $&$ -N_{\ev}$&$ -N_{\ev} $  \\
\hline
\end{tabular}
\end{center}

\begin{center}
\begin{tabular}{|c|c|c|c|c|c|c|c|c|}
\hline
odd sector & $\Gamma_{1}^o$ & $\Gamma_2^o$  & $S_o(0)$ & $w_1^o$
& $w_2^o$ & $w(q(\Om_o)$ \\ \hline\hline
g.c.\ &$ 1 $&$ S_{\odd} $& $1$ & $ 0 $&$
-N_{\odd}$&$ -N_{\odd} $  \\\hline
e.c.\ &$ r_{\odd} $&$ S_{\odd} $&$-1$&$  \frac{1}{2} $&$
-(N_{\odd}+\frac{1}{2}) $&$ -N_{\odd} $  \\\hline
\end{tabular}
\end{center}
Summing up the results of both sectors one obtains \eqref{presquefini}
as there is never an even and an odd zero energy resonance at the same time.

\section{Conclusion}

Levinson's theorem is an index theorem.
We have elaborated the general framework supporting this statement,
and corroborated it with one-dimensional scattering systems with point
interaction or sufficiently fast decreasing potentials.
Our formulation reveals its topological nature and explains
the corrections in a coherent and natural way.
The proof is based on a new formula for the wave-operator
involving up to a compact operator the scattering operator and a
universal function of the dilation operator. This formula is of independent
interest and might be of use in other contexts as well.

\end{document}